\documentclass[submission,copyright,creativecommons]{eptcs}
\providecommand{\event}{ICLP 2022} 

\usepackage{breakurl}             
\usepackage{underscore}           
\usepackage{amsmath}
\usepackage{graphicx}
\usepackage{epsf}
\usepackage{algorithm}
\usepackage[noend]{algpseudocode}
\usepackage{algpseudocode}
\usepackage{graphics}
\usepackage{multirow}
\usepackage{caption}
\usepackage{subcaption}
\usepackage{tabularx}
\usepackage{url}

\title{Knowledge Authoring with Factual English\thanks{Research partially funded by the NSF Grant 1814457. We would also like to thank Nathanael Payen for his contribution to software development for this work.}}
\author{Yuheng Wang \qquad Giorgian Borca-Tasciuc \qquad Nikhil Goel \\ Paul Fodor \qquad Michael Kifer
\institute{Department of Computer Science, Stony Brook University}
\institute{Stony Brook, NY, USA}
\email{\{yuhewang,gborcatasciu,nigoel,pfodor,kifer\}@cs.stonybrook.edu}
}
\def\titlerunning{Knowledge Authoring with Factual English}
\def\authorrunning{Y. Wang, G. Borca-Tasciuc, N. Goel, P. Fodor, \& M. Kifer}
\def\copyrightholders{Wang, Borca-Tasciuc, Goel, Fodor, \& Kifer}

\newcommand\code[1]{\texttt{#1}}
\newcommand\ud[1]{\texttt{#1$^{\texttt{UD}}$}}
\newcommand\xpos[1]{\texttt{#1$^{\texttt{X}}$}}
\newcommand\upos[1]{\texttt{#1$^{\texttt{U}}$}}
\newcommand\KALMF{{KALM\textsuperscript{FL}}}
\newcommand\Stanza{{\textsc{Stanza}}}
\newcommand\MS{{m\textsc{Stanza}}}

\newcounter{example}
\newenvironment{example}{\refstepcounter{example}\noindent\textbf{Example~\theexample}}

\newcounter{property}
\newenvironment{property}{\refstepcounter{property}\noindent\textbf{Property~\theproperty}}

\newcounter{definition}
\newenvironment{definition}{\refstepcounter{definition}\noindent\textbf{Definition~\thedefinition}}

\begin{document}
\maketitle

\begin{abstract}
Knowledge representation and reasoning (KRR) systems represent knowledge as collections of facts and rules. Like databases, KRR systems contain information about domains of human activities like industrial enterprises, science, and business.
KRRs can represent complex concepts and relations, and they can query and manipulate information in sophisticated ways.
Unfortunately, the KRR technology has been hindered by the fact that specifying the requisite knowledge requires skills that most domain experts do not have, and professional  knowledge engineers are hard to find.
One solution could be to extract knowledge from English text, and a number of works have attempted to do so (OpenSesame, Google's Sling, etc.).
Unfortunately, at present, extraction of logical facts from unrestricted natural language is still too inaccurate to be used for reasoning, while restricting the grammar of the language (so-called controlled natural language, or CNL) is hard for the users to learn and use.
Nevertheless, some recent CNL-based approaches, such as the Knowledge Authoring Logic Machine (KALM),
have shown to have very high accuracy compared to others, and a natural question is to what extent the CNL restrictions can be lifted.
In this paper, we address this issue by transplanting the KALM framework to a neural natural language parser, \MS.
Here we limit our attention to authoring
facts and queries and therefore our focus is what we call \emph{factual} English statements.
Authoring other types of knowledge, such as rules, will be considered in our followup work.
As it turns out, neural network based parsers have problems of their own and the mistakes they make range from part-of-speech tagging to lemmatization to dependency errors. We present a number of techniques for combating these problems and test the new system, \KALMF~(i.e., KALM for factual language), on a number of benchmarks, which show \KALMF~achieves correctness in excess of 95\%.
\end{abstract}


\section{Introduction}

Much of the human knowledge can be captured in knowledge representation and reasoning (KRR) systems that are based on logical facts and rules. Unfortunately, translating human knowledge into the logic form that can be used by KRR systems requires well-trained domain experts who are hard to come by.

One popular idea is to use natural language (NL) to represent knowledge, but current technology (e.g. OpenSesame~\cite{swayamdipta2017frame}, SLING~\cite{ringgaard2017sling}) for converting such statements into logic has rather low accuracy.
A possible fix to this problem is to author knowledge via sentences in
controlled natural languages (CNLs), such as ACE used in Attempto~\cite{fuchs1996attempto}. These CNLs are fairly rich and algorithms exist for converting CNL sentences into logic facts.
Unfortunately, CNLs are also very restrictive, hard to extend, and require significant training to use.
Furthermore, both CNLs and the more general NLP systems cannot recognize sentences with identical meaning but different syntactic forms. For example, ``\textit{Mary buys a car}'' and ``\textit{Mary makes a purchase of a car}'' would be translated into totally different logical representations by most systems, which renders logical inference mechanisms unreliable at best. This problem is known as \emph{semantic mismatch}~\cite{gao2018high}.

Recently, the Knowledge Authoring Logic Machine (KALM)~\cite{gao2018knowledge} was introduced to solve the above semantic mismatch problem, but KALM was based on Attempto's ACE and therefore inherited all the aforesaid problems with CNLs.
In this paper, we address the problems associated with controlled languages
by transplanting the KALM framework to
a neural NL parser, \MS, which is a modified \Stanza~\cite{qi2020stanza} version with multiple, ranked outputs.  Of course, to turn English into an authoring tool for KR one still needs to impose some restrictions on the language. For instance, ``Go fetch more water'' is a command that does not convey any factual information that can be recorded in a knowledge base (except, perhaps, those based on rather esoteric logics). In this paper, we
focus on English sentences suitable for expressing facts and queries and correspondingly identify 
a class of English sentences, which we call \emph{factual}. These sentences can be translated into logic and the aforesaid semantic mismatch problem is solved for such sentences.
Unlike CNLs, factual sentences need little training as long as the author keeps focus on knowledge representation rather than fine letters.\footnote{Our followup work will consider more general sentences, such as those suitable for expressing rules.}

To increase the accuracy, we had to mitigate a slew of issues that are common to neural parsers, and we describe our solutions. These include the mistakes in part-of-speech and dependency parsing.
The new system, \KALMF~(KALM for factual language)\footnote{\url{https://github.com/yuhengwang1/kalm-fl}}, is tested on a number of benchmarks, which show that \KALMF~for factual English achieves correctness in excess of 95\%, very close to the original KALM for the Attempto's CNL.

The paper is organized as follows: Section~\ref{sec:kalm} reviews the KALM framework, Section~\ref{sec:fsent} defines factual sentences, Section~\ref{sec:kalmf} proposes \MS~and the new \KALMF~framework, Section \ref{sec:eval} shows the evaluation settings and results, Section~\ref{sec:conclusion} concludes the paper.

\section{The KALM Framework}\label{sec:kalm}

\begin{figure}[htbp!]
    \centering
    \includegraphics[scale=0.3]{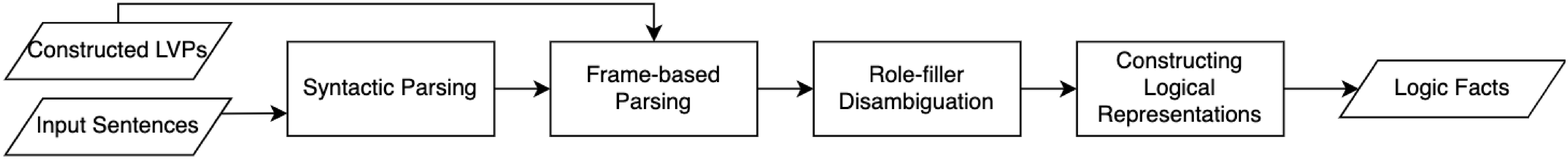}
    \caption{The KALM framework}
    \label{fig:kalm}
\end{figure}

KALM~\cite{gao2018high} is a semantic framework for scalable knowledge authoring. KALM users author knowledge using CNL sentences (Attempto's ACE, to be specific) and KALM ensures that semantically equivalent sentences have identical logical representations through the use of the frame semantics~\cite{fillmore2006frame}. The framework is depicted in Fig.~\ref{fig:kalm}.

\textbf{Syntactic Parsing.} KALM uses Attempto Parsing Engine (APE) to extract the syntactical information from sentences, including the part-of-speech (POS) for each word and the grammatical dependency relations between pairs of words. All extracted information is represented by a set of logical terms known as Discourse Representation Structure (DRS)~\cite{fuchs2006discourse}. Here is an example of a DRS.

\begin{example}
\label{exmp:1}
DRS for the sentence ``\textit{Mary buys a car}''.

{\footnotesize
\begin{verbatim}
    object(A,mary,uncountable,na,eq,1)-1/1.    predicate(C,buy,A,B)-1/2.
    object(B,car,countable,na,eq,1)-1/4.
\end{verbatim}}
\end{example}

\noindent
where \code{A} and \code{B} are identifiers for the \textit{Mary}- and \textit{car}-entities, respectively, and \code{C} is the \textit{buy}-event. The DRS also relies on the predicates \code{object/6} and \code{predicate/4} (in \code{p/N}, \code{N} denotes the number of arguments in predicate \code{p}). An \code{object}-fact represents an entity---a noun-word with some properties (e.g., countable or uncountable, quantity). A \code{predicate}-fact represents an event---a verb-word and its participating entities. Additional predicates, \code{property/3}, \code{relation/3}, \code{modifier\_adv/3}, \code{modifier\_pp/3}, and \code{has\_part/2}, are also used in DRS for representing other syntactic relations. An expression like \code{1/4} in a DRS \code{object}-fact indicates the sentence Id (i.e., 1) and the token Id (here 4) described by the object fact in question.

\textbf{Logical Valence Pattern.} Frame-based parsing requires logical valence patterns (LVPs), which are constructed from \emph{training} sentences annotated by knowledge engineers. Inspired by FrameNet~\cite{baker1998berkeley}, the semantics of each training sentence is represented by a \emph{frame}. Each frame defines one or more related semantic relationships among entities, where each entity plays a particular \emph{role}. A frame can be triggered by its associated ``triggering words,'' called \emph{lexical units (LU)}. Below is an example of a training sentence annotated with frame semantics.

\begin{example}
\label{exmp:2}
The training sentence ``\textit{Mary buys a car}''.

{\footnotesize
\begin{verbatim}
    train('Mary buys a car','Commerce_buy','LUIndex'=2,
          ['Buyer'=1+required,'Goods'=4+required],[[purchase,verb],[acquire,verb]]).
\end{verbatim}}
\end{example}

\noindent
The above says several things.
(i) ``\textit{Mary buys a car}'' is represented by the \code{Commerce\_buy}
frame,
(ii) the LU is the 2nd word, \code{buy.verb}, (iii) the 1st word \textit{Mary} fills the role \code{Buyer} and the 4th word \textit{car} fills the role \code{Goods}, and (iv) the words \code{purchase.verb} and \code{acquire.verb} can also trigger this frame.
Combining this and the DRS in Example~\ref{exmp:1}, KALM learns that starting from the LU \code{buy.verb} (i.e., \code{predicate(C,buy,A,B)}), the \code{Buyer} \textit{Mary} (i.e., \code{object(A,mary,uncountable,na,eq,1)}) can be found by locating the subject (i.e., the 3rd argument in \code{predicate/4}) of \code{buy.verb}.
As a result, the pattern \code{verb->subject} for finding the role-filler for \code{Buyer} is learned.
Similarly, the pattern \code{verb->object} to find \code{Goods} can be learned.
This allows us to construct the following LVP,
where the first 3 arguments identify the LU, the POS of the LU, and the frame.

{\footnotesize
\begin{verbatim}
    lvp(buy,verb,'Commerce_buy',[pattern('Buyer','verb->subject',required),
                                 pattern('Goods','verb->object',required)]).
\end{verbatim}}

\textbf{Frame-based Parsing.} Once the LVPs are constructed, they can be used to extract logical relations from sentences. Namely, when a new sentence comes in, KALM tries to find the triggered LVPs which are then applied to the sentence to get candidate parses.

\begin{example}
\label{exmp:3}
Consider the sentence ``\textit{A customer buys a watch},'' whose DRS is as follows:

{\footnotesize
\begin{verbatim}
    object(A,customer,countable,na,eq,1)-1/2.   predicate(C,buy,A,B)-1/3.
    object(B,watch,countable,na,eq,1)-1/5.
\end{verbatim}}
\end{example}

\noindent
The word \code{buy.verb} triggers the LVP in Example~\ref{exmp:2}. Following the pattern \code{verb->subject} that extracts the roll-filler of \code{Buyer}, KALM starts from the LU \code{buy.verb} (i.e., \code{predicate(C,buy,A,B)}), and then finds the subject of the LU, which is the 3rd argument of \code{predicate(C,buy,A,B)} (i.e., the identifier \code{A}). Finally, the word identified by \code{A} (i.e., \textit{customer}) is extracted as the role-filler for \code{Buyer}. In this way, KALM applies all patterns to all extract role-fillers and finally we have the following candidate parse:

{\footnotesize
\begin{verbatim}
    p('Commerce_buy',[role('Buyer','customer'),role('Goods','watch')]).
\end{verbatim}}

\textbf{Role-Filler Disambiguation.}
Generally, a word is associated with several meanings. The goal of role-filler disambiguation is to find the most appropriate sense for each role-filler with respect to the roles in particular logical frames.
Role-filler disambiguation is done via a walk through the BabelNet knowledge graph~\cite{navigli2012babelnet}. BabelNet combines the words with similar meanings into synset nodes. Edges represent semantic relations (hypernym, hyponym, etc.) and strength of the different relationships is specified via weights.

Consider the candidate parse of ``\textit{A customer buys a watch}'' in Example~\ref{exmp:3}. In BabelNet, the role-filler \textit{watch} has several meanings like ``\textit{A small portable timepiece}'' (\code{bn:00077172n}), ``\textit{A period of time (4 or 2 hours) during which some of a ship's crew are on duty}'' (\code{bn:00080550n}), and more.
Since \code{Goods} (with the synset \code{bn:00021045n}) is much more semantically related to a timepiece than to a period of time, \textit{watch} should be disambiguated with the synset \code{bn:00077172n} denoting a timepiece. Starting from \code{Goods}'s synset \code{bn:00021045n}, KALM uses breadth first search to reach \textit{watch}'s synsets \code{bn:00077172n} and \code{bn:00080550n} respectively, computes the costs based on the edge weights, and ends up with the synset that has the lowest cost, which is \code{bn:00077172n} (``\textit{A small portable timepiece}'') in this case.

\textbf{Constructing Logic Representation.} Ultimately, the disambiguated candidate parses are translated into \emph{unique logical representation} (ULR), which gives the true meaning to the original CNL sentence and is suitable for querying. ULR uses the predicates \code{frame/2} and \code{role/2} for representing instances of the frames and the roles. The predicates \code{synset/2} and \code{text/2} are used to specify synset and textual information. For example, ``\textit{A customer buys a watch}'' will be converted into the ULR shown below:

{\footnotesize
\begin{verbatim}
    frame(id_1,'Commerce_buy').
    role(id_1,'Buyer',id_2). synset(id_2,'bn:00019763n'). text(id_2,'customer').
    role(id_1,'Goods',id_3). synset(id_3,'bn:00077172n'). text(id_3,'watch').
\end{verbatim}}

\section{Factual Sentences}\label{sec:fsent}

In knowledge authoring, we are not interested in fine letters but rather in sentences that express or query knowledge, such as facts, queries, rules, modalities.
In this paper we limit ourselves to facts and queries and more advanced types of knowledge is left to followup papers.
Consequently, here we focus on sentences for specifying and querying sets of facts, which we call \emph{factual sentences}.
Non-factual sentences, like ``\textit{Go fetch more water},'' do not express any factual information and can be thus excluded from consideration.

Before defining factual sentences, we first remind some key grammatical concepts. A \emph{clause} is a unit of grammatical organization that contains a verb and usually other components. A \emph{main clause}\footnote{\url{https://www.lexico.com/en/definition/main_clause}} is a clause that can form a complete sentence standing alone and having a subject and a predicate.
A \emph{subordinate clause} depends on a main clause for its meaning. Together with the main clause, a subordinate clause forms part of a \emph{complex sentence}. There are 4 types of subordinate clauses including adnominal clauses, adverbial clauses, clausal complements and clausal subjects.
A \emph{coordination} is a syntactic structure that links together two or more elements with connectives such as ``\textit{and}'' and ``\textit{or}'' (e.g., \textit{a car and a watch}).
When the elements are main clauses, a \emph{compound sentence} is formed (e.g., ``\textit{Mary wants the car and the car is available}'').

Examination of various datasets shows that main clauses, compound sentences, and sentences with adnominal clauses (e.g. ``\textit{Mary bought a car made in USA}'') are by far the most common constructs in datasets that contain data and queries. In contrast, clausal complements and other types of subordinate clauses are typically non-factual or they are used to describe other kinds of logical statements, such as rules, which will be the subject of our followup work. For the same reason, connectives other than ``\textit{and}" and ``\emph{or}" are also eliminated. We then define \emph{factual sentence} for knowledge authoring as follows:

    
    
        

\begin{definition}
A \emph{factual sentence} is
\begin{enumerate}
    \item a factual main clause with subordinate adnominal clauses (if any), and no other subordinate clauses; or
    \item a compound sentence where the connectives connect only the clauses of the kind described in 1.
\end{enumerate}
\end{definition}

\begin{definition}
A \emph{main clause is factual}\footnote{According to \url{https://www.lexico.com/en/definition/main_clause}, all main clauses are factual.} if
\begin{itemize}
    \item it has a verb with a subject (e.g., ``\textit{Mary bought a car}''); or
    \item it has a nominal word (or an adjective) with a subject and a linking verb (e.g., ``\textit{Mary is rich},'' where an adjective \textit{rich} has a subject \textit{Mary} and linking verb \textit{is})
\end{itemize}
\end{definition}

\subsection{Grammatical Properties of Factual Sentences}\label{sec:prop}
We now use POS tags (part of speech) and universal dependencies to describe six \emph{grammatical properties for factual sentences} that follow from the aforesaid factual restriction on the English sentences, and thus are \emph{necessary conditions} for sentences to be factual.
We then use these properties to discover and correct errors made by the \Stanza~parser.
We use the superscripts \code{U} and \code{X} to refer to universal POS\footnote{\url{https://universaldependencies.org/u/pos/}} (UPOS) tags, Penn Treebank extended POS\footnote{\url{https://www.ling.upenn.edu/courses/Fall_2003/ling001/penn_treebank_pos.html}} tags (XPOS), and \code{UD} will refer to universal dependency\footnote{\url{https://universaldependencies.org/u/dep/}} labels.

\begin{property}
\label{prop:1}
If the main clause is factual, then
    \begin{itemize}
        \item the main clause has a verb with a subject. That is, the clause has a word with an incoming \ud{root} edge tagged with \upos{VERB} and an outgoing \ud{nsubj} edge); or
        \item the main clause has a nominal word (or an adjective) with a subject and a linking verb. Thus, the clause has a word with an incoming \ud{root} edge that is (i) tagged with \upos{NOUN}, \upos{PRON}, \upos{PROPN},  or \upos{ADJ}; (ii) has an outgoing \ud{nsubj} edge; and (iii) has an outgoing \ud{cop} edge (copula).
    \end{itemize}
\end{property}

\begin{property}
\label{prop:6}
If a word $W$ is the last element of a coordination (e.g., ``\textit{watch}" in ``\textit{a car or a watch}"), then this coordination must be an \textit{and}- or an \textit{or}-coordination. That is, $W$ has an incoming \ud{conj} edge and a outgoing \ud{cc} edge pointing to ``\textit{and}'' or ``\textit{or}.''
\end{property}
    
\begin{property}
\label{prop:2}
If a verb $V$ has one or more auxiliary verbs ${V_1^a,..., V_n^a}$, and $V_n^a$ (tagged with \upos{AUX}) is the closest auxiliary verb to $V$ (e.g., in the sentence ``\textit{A car has been bought by Mary}," $V_1^a=$ \textit{has}, $V_n^a=V_2^a=$ \textit{been}, $V=$ \textit{bought}), then
    \begin{itemize}
        \item continuous tense ($V_n^a$ is \textit{be} -- $V$ is a present participle): $V_n^a$ has an incoming \ud{aux} edge starting from $V$, and $V$ is tagged with \xpos{VBG}; or
        \item perfect tense ($V_n^a$ is \textit{have} -- $V$ is a past participle): $V_n^a$ has an incoming \ud{aux} edge starting from $V$, and $V$ is tagged with \xpos{VBN}; or
        \item past, present, and future tense ($V_n^a$ is \textit{can/do/may/must/ought/should/will} -- $V$ in base form): $V_n^a$ has an incoming \ud{aux} starting at $V$, and $V$ is tagged with \xpos{VB}; or
        \item passive voice ($V_n^a$ is \textit{be/get} -- $V$ is a past participle): $V_n^a$ has an incoming \ud{aux:pass} edge starting from $V$, and $V$ is tagged with \xpos{VBN}
    \end{itemize}
\end{property}

\begin{property}\label{prop:3}
For a verb $V$ without auxiliary verbs (no outgoing \ud{aux}/\ud{aux:pass} edges):
    \begin{enumerate}
    
        \item if $V$ is a present or past participle (i.e., tagged with \xpos{VBG}/\xpos{VBN}), then
        \begin{itemize}
            \item $V$ occurs in a coordination (i.e., has an incoming \ud{conj} edge); or
            \item $V$ occurs in adnominal clauses (i.e., has an incoming \ud{acl} edge)
        \end{itemize}

        \item if $V$ is in present or past tense (i.e., tagged with \xpos{VBP}/\xpos{VBZ}/\xpos{VBD}), then
            \begin{itemize}
                \item $V$ occurs in a coordination (i.e., has an incoming \ud{conj} edge); or
                \item $V$ occurs in main/adnominal clauses (i.e., has an incoming \ud{root}/\ud{acl}/ \ud{acl:relcl} edge) and have a subject (i.e., an outgoing \ud{nsubj} edge)
            \end{itemize}
        
        \item if $V$ is in the base form (i.e., tagged with \xpos{VB}), then
            \begin{itemize}
                \item $V$ occurs in a coordination (i.e., has an incoming \ud{conj} edge); or
                \item $V$ occurs in adnominal clauses with infinitive form (i.e., has an incoming \ud{acl} edge and an outgoing \ud{mark} edge pointing to ``\textit{to}'')
            \end{itemize}
        
    \end{enumerate}
\end{property}

\begin{property}\label{prop:4}
If a non-verb word $W$ has one or more auxiliary verbs ${V_1^a,..., V_n^a}$, where $V_n^a$ is the closest auxiliary verb to $W$ (e.g., in the sentence ``\textit{Mary has been rich}," $V_1^a=$\textit{has}, $V_2^a=$ \textit{been}," $W=$ \textit{rich}, and \textit{been} is the closest auxiliary verb to \textit{rich}), then $V_n^a$ and $W$ must satisfy these properties:
    \begin{enumerate}
        \item $W$ is a nominal word or an adjective (i.e., tagged with \upos{NOUN}/\upos{PRON}/\upos{PROPN}/\upos{ADJ})
        \item $V_n^a$ is the copula of $W$ (i.e., $V_n^a$ has an incoming \ud{cop} edge starting from $W$)
        \item $W$ has a subject (i.e., has an outgoing \ud{nsubj} edge)
    \end{enumerate}
\end{property}        

\begin{property}\label{prop:5}
The sentence must be \emph{projective}. Given a parse, if there are crossing edges (e.g., the incoming edges for ``\textit{of}'' and ``\textit{Winston}'' in Fig.~\ref{fig:stz20}), the sentence is called \emph{non-projective}, otherwise it is \emph{projective}. Property~\ref{prop:5} expresses the belief held by linguists that well-constructed English sentences are typically projective, and so are factual sentences.
\begin{figure}[htbp!]
    \centering
    \includegraphics[scale=0.42]{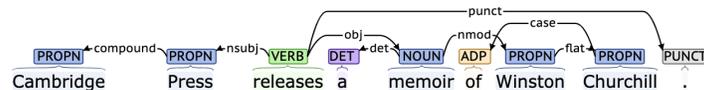}
    \caption{A example of a non-projective parse}
    \label{fig:stz20}
\end{figure}
\end{property}


\section{KALM for Factual Language}\label{sec:kalmf}

\begin{figure}[htbp!]
    \centering
    \includegraphics[scale=0.3]{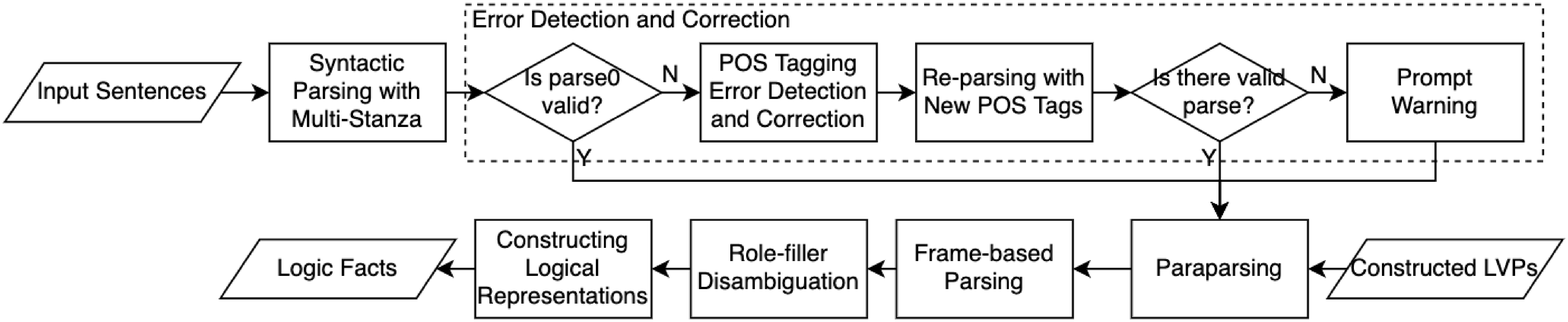}
    \caption{The \KALMF~framework}
    \label{fig:kalmf}
\end{figure}

We now briefly describe a neural parser multi-\Stanza~(\MS) that generates several ranked parses for input sentences and is a modification of the original \Stanza.
Then we describe \KALMF, a product of adaptation of the KALM framework to factual English sentences---a language that is significantly less restricted than any known CNL, and is much easier to learn. 
The \KALMF{} framework is shown in Fig.~\ref{fig:kalmf}.

\begin{figure}[htbp!]
    \centering
    \includegraphics[scale=0.3]{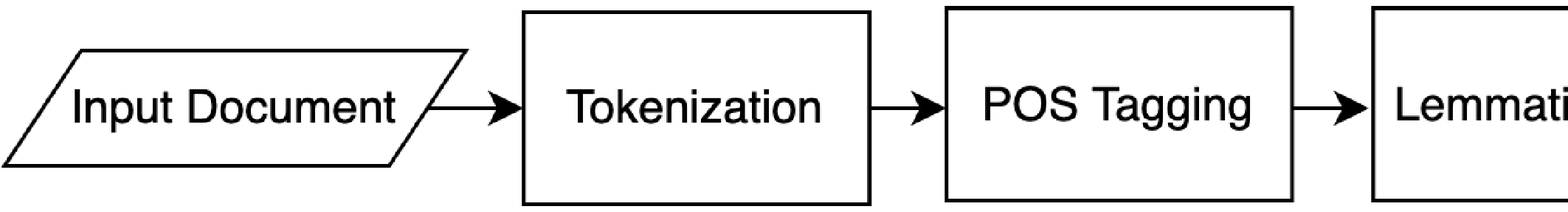}
    \caption{The architecture of \MS}
    \label{fig:ms-arch}
\end{figure}

\subsection{Multi-\Stanza}\label{sec:ms}
\Stanza~\cite{qi2020stanza} is a pipelined neural parser with state-of-the-art performance, which was designed to return only the top parse for each sentence.
Unfortunately, we found that it frequently errs in POS tagging, and these errors then propagate to universal dependencies. We then noticed that nearly top parses often give correct POS tags where the top parses err and so we modified \Stanza~to return also some non-top-ranked parses. We called the result \MS.
Figure~\ref{fig:ms-arch} shows the architecture.
Each stage adds multiple sets of annotations, creating a new \texttt{Document} object for each set. These \texttt{Document} objects are then passed downstream.
Unlike \Stanza, the output of \MS~is a list of annotated \texttt{Document} objects ranked in the order of decreasing confidence.



\subsubsection{POS Tagger and Dependency Parser}
The Part-of-Speech (POS) tagger adds POS tags to each word. As each POS tag has a finite number of categories, it is straightforward to extract the $k$-best POS tags for each word, along with their confidences. \MS~allows a user to provide a function to dynamically modify the list of $k$-best POS tags according to their needs. In terms of dependency parsing, \MS~generates a dependency parse by generating a fully connected directed graph, and generating the weights of the edges using a neural network.
The neural network learns to assign the weight of the edge based on the type of edge and the relationship between the vertices. Then, the minimum spanning arborescence is found and used as the dependency parse. Multiple possible dependency parses for each sentence are combined in order to generate the next-best parse for the entire document.


\subsubsection{Error Detection and Correction Based on Multi-\Stanza}

\KALMF~checks \Stanza~parses for being factual using the necessary conditions of Section~\ref{sec:prop}. If any of the checks don't pass, \KALMF~attempts to correct the parse by conjecturing that some of the POS tags are wrong (a fairly common problem with \Stanza~in our experience). This is done by using other nearly top parses provided by \MS. If the correction attempt fails, the user is asked to rephrase the sentence. Details of the error correction algorithm are given in
\ref{appdx:correct}.

\subsection{Paraparsing}\label{sec:paraparsing}

Paraparsing is a set of corrective steps that modify the original $\overline{parse}$ (see Appendix \ref{appdx:correct}). The aim here is to
eliminate possible semantic mismatches that were the original motivation for KALM, as explained in the introduction. The mismatches handled here arise from the possibility that the same information may be described via passive or active voice, via a different order of elements in a coordination, via the different ways to attach adnominal clauses, and more. Note that all these corrections became possible in \KALMF{} due to
the use of dependency parsing and were not possible in the original CNL-based KALM.

\subsubsection{Passive Voice}
\begin{figure}[htbp!]
     \centering
     \begin{subfigure}[b]{0.3\textwidth}
         \centering
         \includegraphics[width=\textwidth]{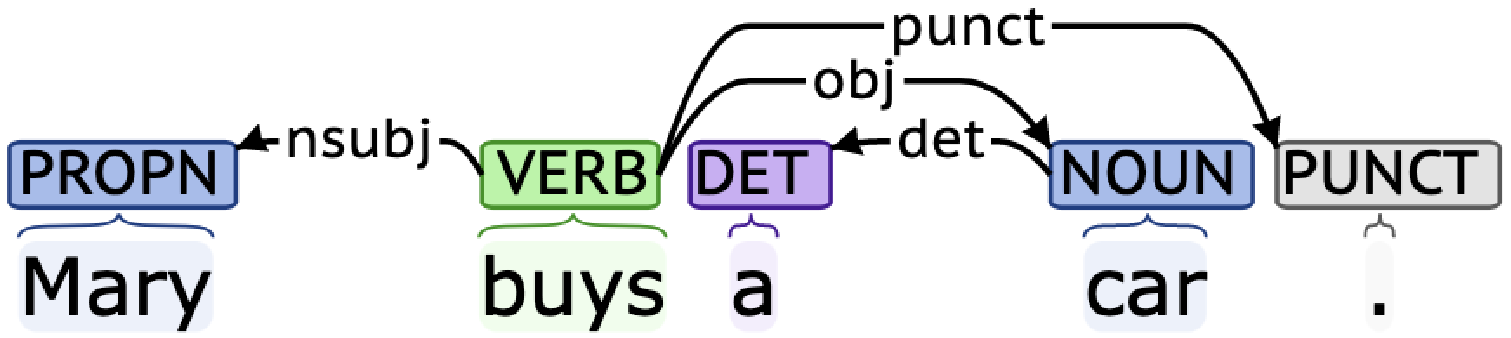}
         \caption{Active voice}
         \label{fig:stz2}
     \end{subfigure}
     \hspace{5mm}
     \begin{subfigure}[b]{0.43\textwidth}
         \centering
         \includegraphics[width=\textwidth]{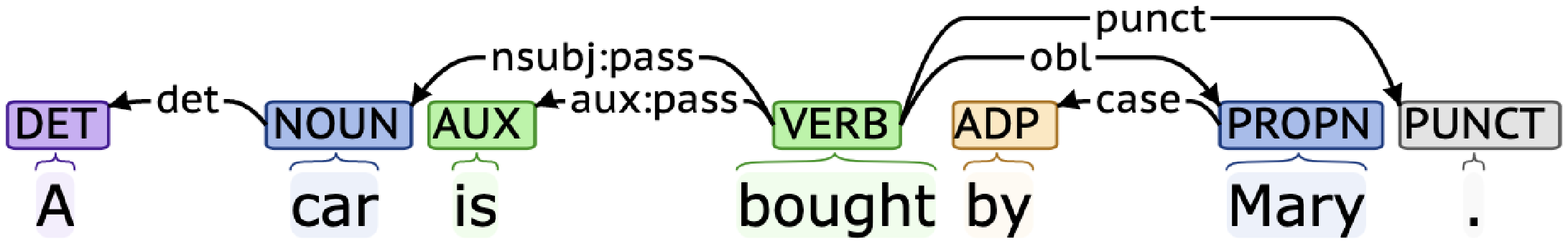}
         \caption{Passive voice}
         \label{fig:stz22}
     \end{subfigure}
     \caption{Semantic mismatch caused by passive voice}
     \label{fig:passive}
\end{figure}

\MS~handles the active and passive voices separately. For a pair of active/passive voice sentences with the same meaning, such as ``\textit{Mary buys a car}'' and ``\textit{A car is bought by Mary},'' \Stanza~gives two completely different parses shown in Fig.~\ref{fig:passive}; it does not attempt to reconcile the semantic mismatch between them so that they would yield the same logical representation.
To address this problem, \KALMF~first recognizes passive voice by the \ud{aux:pass} edge in the parse, then modifies the edges of passive voice parses to make the parses equivalent to their active voice counterparts. If the sentence is in active voice, keep the parse unchanged. Otherwise, convert it into $n$ parses in active voice ($n$ is the number of \textit{by}-phrases in the clause, since every \textit{by}-phrase could be the subject of the real active voice counterpart of this passive voice sentence) by modifying (i) \ud{nsubj:pass} to \code{obl:by} and (ii) $n$ \code{obl:by} edges to \ud{nsubj} one by one.

\subsubsection{Coordination}

\begin{figure}[htbp!]
     \centering
     \begin{subfigure}[b]{0.49\textwidth}
         \centering
         \includegraphics[width=\textwidth]{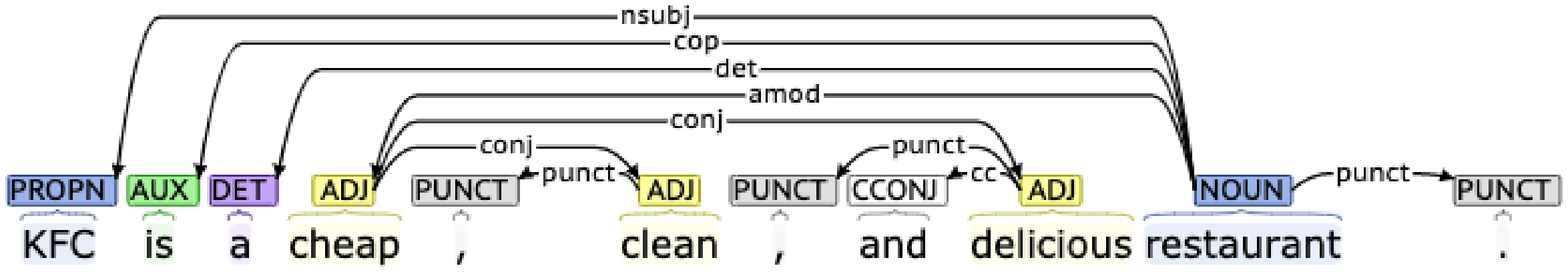}
         \caption{}
         \label{fig:stz5}
     \end{subfigure}
     \begin{subfigure}[b]{0.49\textwidth}
         \centering
         \includegraphics[width=\textwidth]{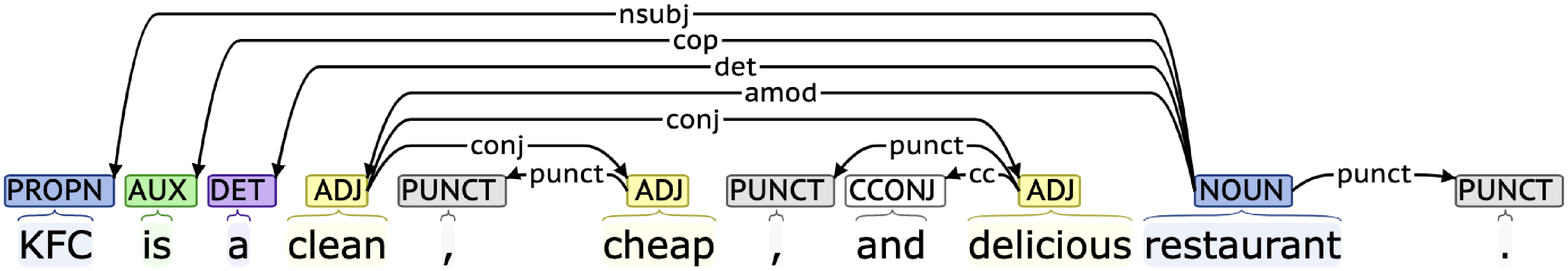}
         \caption{}
         \label{fig:stz35}
     \end{subfigure}
     \caption{Semantic mismatch caused by coordination}
\end{figure}

Elements in \Stanza~coordinations are not treated equally. For example, in the parse of ``\textit{KFC is a cheap, clean, and delicious restaurant}'' shown in Fig.~\ref{fig:stz5}, \textit{cheap} directly depends on \textit{restaurant}, but \textit{clean} and \textit{delicious} mutually depend on \textit{cheap} instead of \textit{restaurant}. In this case, if \textit{cheap} and \textit{clean} are swapped, the meaning of the sentence stays unchanged, but the parse will be different as shown in Fig.~\ref{fig:stz35}. This phenomenon will lead to a semantic mismatch.

\KALMF~treats coordination elements equally by modifying their edges. The procedure is shown below, and all examples refer to Fig.~\ref{fig:stz5}.
    \begin{enumerate}
        \item Locate the root element $el_{root}$ of the coordination. It is a word that has outgoing \ud{conj} edges to other elements, which have incoming \ud{conj} edges (e.g. $el_{root}$ is \textit{cheap}, while \textit{clean} and \textit{delicious} are the other two elements)
        \item Copy the incoming edge of $el_{root}$ to each non-root element and delete the edge \ud{conj} (e.g. copy \ud{amod} to replace \ud{conj} that goes to \textit{clean} and \textit{delicious})
        \item Copy the outgoing edges of $el_{root}$ (other than the deleted \ud{conj}) to each non-root element (in our example, \textit{cheap} has no outgoing edges, so no need to copy anything)
    \end{enumerate}

\subsubsection{Adnominal Clause}
\begin{figure}[htbp!]
     \centering
     \begin{subfigure}[b]{0.585\textwidth}
         \centering
         \includegraphics[width=\textwidth]{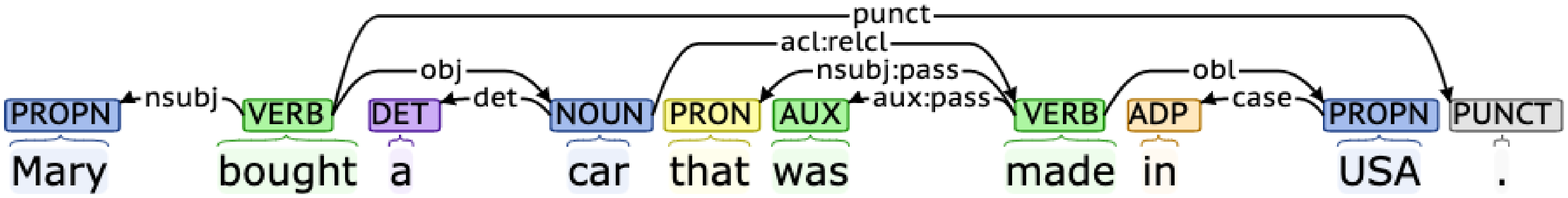}
         \caption{}
         \label{fig:stz37}
     \end{subfigure}
     \begin{subfigure}[b]{0.405\textwidth}
         \centering
         \includegraphics[width=\textwidth]{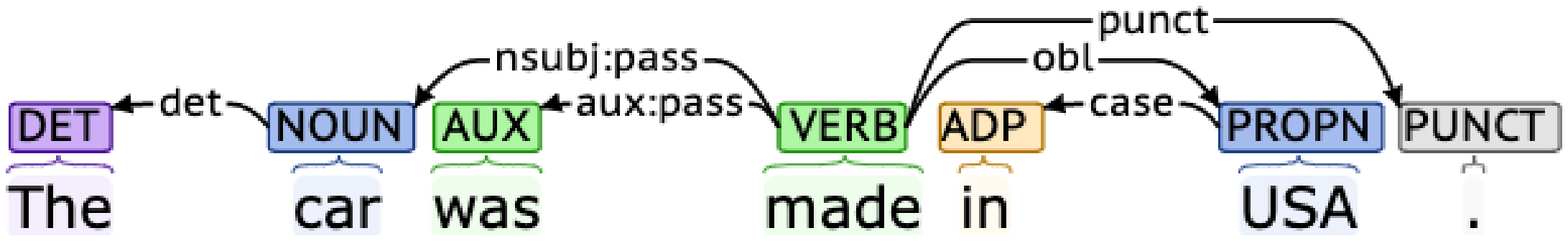}
         \caption{}
         \label{fig:stz38}
     \end{subfigure}
     \caption{Semantic mismatch caused by adnominal clause}
\end{figure}

An adnominal clause describes a fact about the nominal word it modifies. For example, ``\textit{Mary bought a car that was made in USA}'' represents two facts: ``Mary bought a car,'' and ``The car was made in USA''. However, the second fact is parsed differently when it is in an adnominal clause than when it is in a sentence by itself, and such phenomena lead to semantic mismatch. As shown in Fig.~\ref{fig:stz37}, the subject of ``\textit{a car that was made in USA}'' is ``\textit{that}'' whereas the real subject should be ``\textit{car}'' like the parse in Fig.~\ref{fig:stz38}.

\KALMF~recognizes the real roles (e.g., subject, object, etc.) that the modified word plays in the adnominal clause, so that the adnominal clause can be seen as a complete sentence all by itself.
This is done via the following transformation.

\begin{itemize}
    \item if a word $V_1$ has an incoming \ud{acl} edge $e_1$ that starts at $V_2$ and has no outgoing \ud{nsubj} or \ud{nsubj:pass} edges, then 
        \begin{itemize}
            \item if $V_1$ is a present participle or a base-form verb tagged with \xpos{VBG} or \xpos{VB}, flip the direction of $e_1$ and change the label to \ud{nsubj}
            \item if $V_1$ is a past participle tagged with \xpos{VBN}, flip the direction of $e_1$ and change the label to \ud{nsubj:pass}
        \end{itemize}

    \item if a word $V_1$ has an incoming \ud{acl:relcl} edge $e_1$ that starts at $V_2$, then
        \begin{itemize}
            \item if $V_1$ has an outgoing \ud{nsubj}, \ud{nsubj:pass} or \ud{obj} edge pointing to an introductory word $W_{intro}$ ``\textit{that}/\textit{who}/\textit{which},'' replace $W_{intro}$ with $V_2$
            \item if $V_1$ has an outgoing \ud{mark} edge pointing to an introductory word $W_{intro}$ \textit{where}/\textit{when}/\textit{why}/\\\textit{which} replace $W_{intro}$ with $V_2$ and modify \ud{mark} to \ud{obl}
        \end{itemize}
    
\end{itemize}

\subsubsection{Other Semantic Mismatches}

Besides the most frequent semantic mismatch issues solved above, \KALMF~also tackles other types of semantic mismatch caused by lemmatization, particle verbs, prepositional phrases, named entities, indirect objects, and so forth.
As shown in Fig.~\ref{fig:kalmf}, after the Paraparsing step, the ultimate parse is delivered to Frame-based Parsing and undergoes further processing to ultimately yield a unique disambiguated logical representation.

\subsection{Representing Dependencies in Logic Programming Systems}

The original KALM used DRS to represent extracted information. In this paper, the parses are represented by much more general graphs, so we introduce an appropriate logical representation for them. Here is an example shown below:


\begin{example}
\label{exmp:4}
The \KALMF{} representation for the sentence ``\textit{Mary buys a car}''.

{\footnotesize
\begin{verbatim}
token(index(1,1,1),mary,[edge(index(1,2),jbusn)],edge(index(1,2),nsubj),
      propn,nnp,index(1,2),s_person,accepted).
token(index(1,2,1),buy,[edge(index(1,1),nsubj),edge(index(1,4),obj)],edge(index(1,0),root),
      verb,vbz,index(1,2),o,accepted).
token(index(1,3,1),a,[edge(index(1,4),ted)],edge(index(1,4),det),
      det,dt,index(1,2),o,accepted).
token(index(1,4,1),car,[edge(index(1,3),det),edge(index(1,2),jbo)],edge(index(1,2),obj),
      noun,nn,index(1,2),o,accepted).

\end{verbatim}}
\end{example}
\noindent
where a sentence is represented by a set of \code{token/9} predicates and each \code{token/9} predicate represents a token $t$. The 1st argument in a \code{token/9} predicate includes sentence ID, candidate parse ID, and $t$'s ID. The 2nd argument is the lemma of $t$. The 3rd argument is a list of edges that connect $t$ to other tokens, and an \code{edge/2} predicate representing a specific edge $e$ includes the index of the other token on $e$, and the edge type (reversed if it is an in-coming one). The 4th argument is the one and only in-coming edge that $t$ has. The 5th, 6th are $t$'s UPOS and XPOS tags, respectively. The 7th argument is the index of the root token in the whole sentence, namely, ``\textit{buys}" in Example \ref{exmp:4}. Finally, the 8th and 9th arguments are the named entity and validation tags, where the latter indicates if the parse is factual (\code{accepted}) or not.


\subsection{Role-filler Disambiguation and Unique Logical Representations}

In KALM, a clause represents a complete fact so each clause has only one parse, the one with the highest semantic score after disambiguation.
In \KALMF, coordinations and adnominal clauses are introduced and their meanings can be captured accordingly, which is further explained in Section~\ref{sec:coor} and \ref{sec:adn}.


For logical representations, \KALMF~uses \code{ulr/3} and \code{role/4} for representing instances of final parses after disambiguation. Consider theses sentences: ``\textit{Mary bought a car for John}'' and ``\textit{Mary made a purchase of a car for John}''. Although they have different syntactic structures, they are ultimately converted into exactly the same parse and their role-fillers are assigned exactly the same synsets. Therefore, they must be translated into a unique logical representation (ULR). And indeed, the ULR for sentences ``\textit{Mary bought a car for John}'' and ``\textit{Mary made a purchase of a car for John}'' is the same:

{\footnotesize
\begin{verbatim}
    ulr(fid_1,'Commerce_buy',[role(rid_1,'Buyer',mary,'bn:00046516n'),
                              role(rid_2,'Goods',car,'bn:00007309n'),
                              role(rid_3,'Recipient',john,'bn:00046516n')]).
\end{verbatim}}

\noindent
The first argument \code{fid\_1} is the unique ID of this buying event, the second argument,
\code{'Commerce\_buy'}, is the frame name, and argument 3 is a list of role descriptors.

\subsubsection{ULR for Factual Sentences with Coordinations}\label{sec:coor}


Generally, a sentence can have a mixture of \textit{and}- and \textit{or}-coordinations, whose meaning is quite hard to describe. For simplicity, we focus on the case where $C$ has only one type of coordination, i.e., all connectives are \textit{and} or all are \textit{or}.

Let $[C_1,...,C_n]$ be the list of all coordinations in a sentence $S$. A \emph{coordinated choice} is a list $\sigma=[el_1, ..., el_n]$ of coordination elements such that $el_i\in C_i$ for all $i=1,...,n$.
Let $S_\sigma$ be $S$ where each coordination, $C_i$, and its elements is replaced with the corresponding element $el_i$ from $\sigma$. Thus, for each coordinated choice $\sigma$ for $S$, the above replacement operation constructs another sentence, $S_\sigma$, which has no coordinations. Next, we collect $S_\sigma$ for all the different $\sigma$'s and organize these sentences as elements of a new homogeneous coordination of the same type as each of the original coordination $C_i$. The result is a sentence $S^\prime$ with only one coordination, found at the root of the parse for the sentence.

For example, for the sentence ``\textit{Mary bought and sold a car and a watch}'', the coordinated choices includes $\sigma_1=[bought,car]$, $\sigma_2=[bought,watch]$, $\sigma_3=[sold,car]$, $\sigma_4=[sold,watch]$. Based on the coordinated choices, we can construct 4 sentences without coordinations: ``\textit{Mary bought a car},'' ``\textit{Mary bought a watch},'' ``\textit{Mary sold a car},'' and ``\textit{Mary sold a watch},'' which are organized into a new \textit{and}-coordination. Thus, we have the final ULR for this \textit{and}-coordination shown below:

{\footnotesize
\begin{verbatim}
    ulr(fid_1,'Commerce_buy',[role(rid_1,'Buyer',mary,'bn:00046516n'),
                              role(rid_2,'Goods',car,'bn:00007309n')]).
    ulr(fid_2,'Commerce_buy',[role(rid_1,'Buyer',mary,'bn:00046516n'),
                              role(rid_3,'Goods',watch,'bn:00077172n')]).
    ulr(fid_3,'Commerce_sell',[role(rid_1,'Seller',mary,'bn:00046516n'),
                              role(rid_2,'Goods',car,'bn:00007309n')]).
    ulr(fid_4,'Commerce_sell',[role(rid_1,'Seller',mary,'bn:00046516n'),
                              role(rid_3,'Goods',watch,'bn:00077172n')]).
\end{verbatim}}

\subsubsection{ULR for Factual Sentences with Adnominal Clauses}\label{sec:adn}
An adnominal clause always describes the nominal word it modifies. This means that an adnominal clause expresses a complete fact about the nominal word. In other words, the facts represented by adnominal clauses and by the main clause must both hold. Thus, the ULRs for clauses must be in conjunction.
For example, the sentence ``\textit{[Mary bought a car]\textsubscript{main} [made in the country]\textsubscript{adnominal} [that John lives in]\textsubscript{adnominal}}''
has a main clause and two adnominal clauses, one modifying the word ``\emph{car}'' and the other the word
``\emph{country}.''
The ULR then is given below:

{\footnotesize
\begin{verbatim}
    ulr(fid_1,'Commerce_buy',[role(rid_1,'Buyer',mary,'bn:00046516n'),
                              role(rid_2,'Goods',car,'bn:00007309n')]).
    ulr(fid_2,'Manufacturing',[role(rid_2,'Product',car,'bn:00007309n'),
                               role(rid_3,'Place',country,'bn:00023236n')]).
    ulr(fid_3,'Residence',[role(rid_4,'Resident',john,'bn:00046516n'),
                           role(rid_3,'Location',country,'bn:00023236n')]).
\end{verbatim}}


\section{Evaluation}\label{sec:eval}

We use four datasets to demonstrate the high performance of \KALMF~as a knowledge authoring machine for factual English.

\subsection{Datasets}

\begin{itemize}
\item
\textbf{CNLD.} \cite{gao2018knowledge} uses CNL sentences, largely inspired by FrameNet, to evaluate the original KALM. We call this dataset the CNL Dataset (CNLD). CNLD contains 250 short CNL sentences in present tense, such as ``\textit{Kate purchases a house}.'' This dataset is captured via 50 logical frames and 317 LVPs constructed from 213 training sentences.

\item
\textbf{CNLDM.} This dataset is obtained from CNLD by changing the voice of some sentences from active to passive and vice versa. In addition, some sentences are changed to past or future tense.
Thus, CNLDM contains sentences like ``\textit{A house was purchased by Kate}," with mixed voice and tense. Our evaluation uses the same LVPs as in CNLD.
  
\item
\textbf{MetaQA.} This dataset~\cite{zhang2017variational} has queries that use complex adnominal clauses. These queries neatly fall into several different templates. Within each template, the queries differ only in the entity names.
Also, all named entities are pre-annotated. For example, the queries ``\textit{who directed the movies written by [Thomas Ian Griffith]}'' and ``\textit{who directed the movies written by [Frank De Felitta]}'' belong to the same template ``\textit{who directed the movies written by [MASK]},'' where \textit{[MASK]} is a placeholder for pre-annotated named entities.
In this evaluation, we use 2- and 3-hop templates directly instead of the original queries, because with named entities annotated, different queries that fall into a same template have exactly the same \MS~parse.  Only 3 frames are needed to represent the semantics of all such queries: \code{Movie}, \code{Inequality}, and \code{Coop}(eration). Acting as knowledge engineers, we designed 85 training sentences and used the approach of~\cite{gao2019querying} to understand 2- and 3-hop queries.
  

  
\item
\textbf{NLD.} NLD uses part of the dataset from FrameNet. NLD includes 250 sentences which look like: ``\textit{GDA has purchased the site from Laing Homes and plans are being prepared for an 80 million dollar mixed development for business, media and leisure activities}'', which is the original sentence of the CNLD sentence ``\textit{Kate purchases a house}". NLD sentences have much more complicated structures that goes beyond factual sentences. Besides factual parts, most of the NLD sentences have non-factual parts that are not usable for knowledge acquisition. In view of this, we ignore all the non-factual parts in NLD sentences. Another approach could be highlighting the non-factual parts and letting the user to correct them or eliminate them.
\end{itemize}

For the original KALM, all these datasets have to be manually modified to eliminate future/past tense, 
to put adnominal clauses in a certain canonical form, 
restrict the vocabulary for the controlled natural language parser, 
particle verbs, appositives and compound nouns also had to be manually modified.
In \KALMF{}, all this is done automatically, and therefore, it can handle a much bigger share of natural language sentences.

\subsection{Comparison Systems}
We compare \KALMF{} with the original KALM as well as three other
frame-based parsers: SEMAFOR \cite{das2014frame}, SLING~\cite{ringgaard2017sling}, and OpenSesame~\cite{swayamdipta2017frame}.
SEMAFOR and SLING have been previously shown to be inaccurate in~\cite{gao2018high}, so we will not repeat these findings and instead focus on the recently proposed OpenSesame system. Unlike \KALMF, OpenSesame is a three-staged pipeline involving target (i.e., LU) identification, frame identification and argument (i.e., role-filler) identification---each stage is essentially a neural network trained independently of the others. In addition, we consider the neural system DrKIT~\cite{dhingra2020differentiable} as another comparison system, which achieves the best performance on MetaQA
among neural models.

\subsection{Results}

The evaluation is based on the following metrics:

\begin{enumerate}
    \item \textbf{Frame-level Micro-F1}: 
    the ratio of sentences that (i) correctly trigger all the applicable frames, and (ii) do not trigger wrong frames.
    \item \textbf{Role-level Micro-F1}: the ratio of sentences that (i) correctly trigger all the applicable frames with all roles correctly identified, and (ii) do not trigger wrong frames.
    \item \textbf{Synset-level Micro-F1}: 
    the ratio of sentences that (i) correctly trigger all the applicable frames with all roles correctly identified and disambiguated, and (ii) do not trigger wrong frames.
    Note this metric applies only to KALM and \KALMF; other systems do not attempt to give semantics with this level of precision.
\end{enumerate}

\begin{table}[htbp!]
\caption{Micro-F1 score comparisons on different datasets}
\centering
\begin{tabular}{ccccccccccccc}
\hline 
           & \multicolumn{3}{c}{CNLD}                     & \multicolumn{3}{c}{CNLDM}                    & \multicolumn{3}{c}{MetaQA}          & \multicolumn{3}{c}{NLD}                       \\
           & F             & R             & S             & F             & R             & S             & F             & R             & S    & F             & R             & S \\ 
\hline 
KALM       & 0.99          & 0.99          & 0.97          & --           & --           & --           & \textbf{1.00} & \textbf{1.00} & \textbf{1.00} & --           & --           & --           \\
OpenSesame & 0.61          & 0.17          & --           & 0.59          & 0.11          & --           & 0.49          & 0.00          & --  & 0.56          & 0.12          & --           \\
\KALMF      & \textbf{0.99} & \textbf{0.99} & \textbf{0.97} & \textbf{0.99} & \textbf{0.99} & \textbf{0.97} & 0.95          & 0.95           & 0.95 & \textbf{0.99} & \textbf{0.98} & \textbf{0.95} \\
DrKIT      & -- & -- & -- & -- & -- & -- & --          & 0.876          & -- & -- & -- & -- \\
\hline 
\end{tabular}
\label{tab:results}
\end{table}

\noindent
Results for the different levels of F1 scores are presented in Table~\ref{tab:results}. In the table, F, R, and S refer to the frame, role, and synset-level F1 scores, respectively. Note that the results reported in the literature for DrKIT \cite{dhingra2020differentiable} can be interpreted as pertaining the role-level Micro-F1 metric. 
The results in the table are summarized below.
\begin{enumerate}
    \item CNLD: This dataset has only CNL sentences and \KALMF~achieves the same high F1 scores as the original KALM in all metric levels. OpenSesame's 0.61 frame-level F1 score shows that it has difficulty even to recognize correct frames.
    \item CNLDM: Perturbation of the tenses and voices of CNLD sentences took this dataset
    outside of the Attempto's APE CNL, thus the original
    KALM cannot handle some of the CNLDM sentences even though the meaning of these sentence did not change.
    \item MetaQA: KALM performs perfectly, but only after changing the sentences so they comply with the ACE CNL. In contrast, \KALMF~gets
    the 0.95 synset F1 score even without any preprocessing. OpenSesame fails on MetaQA with 0 role-level F1 score---probably because it was never trained on the movie domain. In the comparison between DrKIT and \KALMF, DrKIT~\cite{dhingra2020differentiable} achieved 0.871 and 0.876 accuracy on 2- and 3-hop query answering respectively. For \KALMF, the 333 out of 350 correctly parsed templates covers 128,784 2-hop queries and 119,923 3-hop queries, which results in 0.962 and 0.933 accuracy on 2- and 3-hop query answering and outperforms DrKIT.
    \item NLD: The original KALM fails since this dataset breaks
    the CNL restrictions on the input language. 
    In contrast, \KALMF{} does well and easily outperforms
    OpenSesame, especially when it comes to handling of adnominal clauses.
\end{enumerate}

It is surprising that OpenSesame's role level F1 scores are extremely low on the three FrameNet-related datasets. Error analysis shows that even for simple CNLD sentences like ``\textit{Mary buys a laptop},'' OpenSesame has hard time extracting all roles. For instance, ``\textit{laptop}" is not extracted as a role-filler for the role \code{Goods}.

\section{Limitations and Future Work}\label{sec:limitation}

Although \KALMF{} has been shown to have high accuracy, limitations still exist. For example,
\begin{itemize}
    \item \KALMF{} accepts various tenses based on Property \ref{prop:2}, but currently this and other temporal information is ignored and is planned for future work. 
    \item \KALMF{} doesn't handle anaphora because the quality of the parses is highly dependent on the quality of the chosen off-the-shelf anaphora resolver.
    \item \KALMF{} treats sentences with quantifiers, like ``\textit{Every pet has an owner}," as facts rather than rules, which points to an issue with the definition of factual sentences.
\end{itemize}

For future work, we plan to address some of the aforesaid problems and to extend \KALMF{} with quantifiers, rules, temporal information, and other advanced features that have direct counterparts in natural languages.

\section{Conclusion}\label{sec:conclusion}

The original KALM \cite{gao2018high,gao2018knowledge,gao2019querying} was proposed as a solution to the problem of semantic mismatch in knowledge authoring using natural languages, but this solution was limited to CNLs, which is a severe limitation both in expressiveness and human training. 
In this paper, we introduced \KALMF,
an NLP system that is not chained by CNL limitations.
The only restriction is that the sentences used for knowledge authoring must be factual, i.e., express factual information as opposed to, say, feelings, allegories, hyperbolas, etc. 
Benchmarking shows that this approach captures the meanings of factual sentences with very high accuracy:
the 0.95 F1 score for both facts and queries.

\bibliographystyle{eptcs}
\bibliography{main}

\newpage

\appendix

\section{Error Detection and Correction}\label{appdx:correct}

\begin{figure}[htbp!]
    \centering
    \includegraphics[scale=0.45]{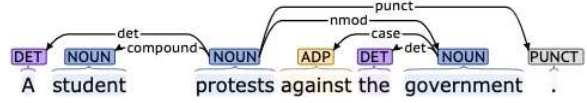}
    \caption{The best \MS~parse for ``\textit{A student protests against the government}''}
    \label{fig:stz3}
\end{figure}

Fig.~\ref{fig:stz3} illustrates one of the errors in \Stanza~POS tagging. Here, the word \textit{protests} is wrongly tagged as a noun and the dependencies related to \textit{protests} are also wrong.

Fortunately, the six properties that stem from the factual sentence requirement
can detect and help correct some of these mis-taggings.
Denote the \MS~parse with the highest confidence score as $parse_0=[w_1,...,w_n]$ where $n$ is the number of the words, $w_i$ contains all parsing information such as POS tag, dependency relation of the $i$-th word in the sentence. Let 
$Upos=[upos_1,...,upos_n]$ and $Xpos=[xpos_1,...,xpos_n]$ be the UPOS and XPOS taggings of the sentence.

\begin{algorithm}[htbp!]
\caption{POS Tagging Error Detection and Correction}
{\bf Input:} $upos_i$, $xpos_i$, the second best UPOS tag $upos_i^\prime$ for $w_i$, and the second best XPOS tag $upos_i^\prime$ for $w_i$.\\
{\bf Output:} The corrected UPOS $\overline{upos_i}$ and XPOS $\overline{xpos_i}$ for $w_i$.
\begin{algorithmic}[1]
\State $\overline{upos_i},\overline{xpos_i} \leftarrow upos_i, xpos_i$
\If{$upos_i.score<0.9$}
    \If {$upos_i==$ \upos{NOUN} and $upos_i^\prime==$ \upos{VERB}}
        \State $\overline{upos_i} \leftarrow$ \upos{VERB}
        \State $\overline{xpos_i} \leftarrow$ \xpos{VBP}/\xpos{VBZ}/\xpos{VBD}
    \ElsIf {$upos_i==$ \upos{VERB} and $upos_i^\prime==$ \upos{AUX}}
        \State $\overline{upos_i} \leftarrow$ \upos{AUX}
        \State $\overline{xpos_i} \leftarrow$ \xpos{VBP}/\xpos{VBZ}/\xpos{VBD}
    \ElsIf {$upos_i==$ \upos{PRON} and $upos_i^\prime==$ \upos{DET}}
        \State $\overline{upos_i} \leftarrow$ \upos{DET}
        \State $\overline{xpos_i} \leftarrow$ \xpos{WDT}/\xpos{PDT}/\xpos{DT}
    \ElsIf {$upos_i==$ \upos{SCONJ} and $upos_i^\prime==$ \upos{ADV}}
        \State $\overline{upos_i} \leftarrow$ \upos{ADV}
        \State $\overline{xpos_i} \leftarrow$ \xpos{WRB}/\xpos{IN}
    \EndIf
\ElsIf{$xpos_i.score<0.9$}
    \If {$xpos_i==$ \xpos{VBD} and $xpos_i^\prime==$ \xpos{VBN}}
        \State $\overline{xpos_i} \leftarrow$ \xpos{VBN}
    \ElsIf{$xpos_i==$ \xpos{VBN} and $xpos_i^\prime==$\ xpos{VBD}}
        \State $\overline{xpos_i} \leftarrow$ \xpos{VBD}
    \ElsIf{$xpos_{1j}==$ \xpos{VBP} and $xpos_i^\prime==$ \xpos{VB}}
        \State $\overline{xpos_i} \leftarrow$ \xpos{VB}
    \EndIf
\EndIf
\State \Return $\overline{upos_i}, \overline{xpos_i}$
\end{algorithmic}
\label{algo:pos}
\end{algorithm}

\textbf{Detection and correction of POS tags.} As shown in the dotted box of Fig.~\ref{fig:kalmf}, this step starts with $parse_0$. If $parse_0$ satisfies all the above-mentioned factual properties, $parse_0$ is assumed to be error-free and is directly sent to the Paraparsing step. Otherwise, following Algorithm~\ref{algo:pos}, if a possibly wrong (with confidence $<$ 0.9) POS tag belongs to a certain type of frequent POS tagging errors (e.g., \MS~relatively frequently mis-tags verbs as nouns, and this is what happened with \textit{protests} in Fig.~\ref{fig:stz3}), \KALMF~then asserts the tag is wrong and corrects it.
Lines 2 and 3 in Algorithm~\ref{algo:pos} capture such type of errors and assign corrected POS tags. Note that in Lines 5, 8, 11, and 14, the algorithm faces multiple options like \xpos{VBP}/\xpos{VBZ}/\xpos{VBD} and chooses the one with the highest confidence score.

\textbf{Re-parsing with new POS tags.} Having re-tagged the words in the above step, the new POS tags,
$\overline{Upos}=[\overline{upos_1},...,\overline{upos_n}]$ and $\overline{Xpos}=[\overline{xpos_1},...,\overline{xpos_n}]$, are fed to the \MS~dependency parser to re-generate a new dependency parse $\overline{Parse}$, ranked by confidence scores.

\textbf{Selecting a corrected parse.} In this step, \KALMF~goes through the parses in $\overline{Parse}$, from the highest confidence score to lowest, looking for a parse, $parse'$, that satisfies all the properties of factual sentences.
If $parse'$ is found, it is taken as a corrected parse, $\overline{parse}=parse'$.
If $parse'$ is not found, the algorithm assumes the sentence in question is not factual, so it asks the user to paraphrase the sentence.

\end{document}